\documentclass[sigconf]{acmart}

\usepackage{verbatim}
\usepackage{booktabs}
\usepackage{tabularx}
\usepackage{subfigure}
\usepackage{enumerate}
\usepackage{enumitem}
\usepackage{xspace}
\usepackage{tikz}
\usepackage{booktabs}

\def\eg{\emph{e.g.}\xspace} 
\def\etc{\emph{etc.}\xspace}
\def\ie{\emph{i.e.}\xspace}
\def\etal{\emph{et al.}\xspace}

\newcommand{\pb}[1]{\vspace{0.75ex}\noindent{\bf \em #1}\hspace*{.3em}}
\newcommand{\pie}[1]{%
    \begin{tikzpicture}
        \draw (0,0) circle (0.75ex);\fill (0.75ex,0) arc (0:(-#1+90):0.75ex) -- (0,0) -- cycle;
        \fill (0.75ex,0) arc (0:(#1-90):0.75ex) -- (0,0) -- cycle;
    \end{tikzpicture}%
}

\AtBeginDocument{%
    \providecommand\BibTeX{{%
        \normalfont B\kern-0.5em{\scshape i\kern-0.25em b}\kern-0.8em\TeX}
    }
}

\copyrightyear{2020}
\acmYear{2020}
\setcopyright{acmcopyright}
\acmConference[EASE 2020]{Evaluation and Assessment in Software Engineering}{April 15--17, 2020}{Trondheim, Norway}
\acmPrice{15.00} \acmDOI{10.1145/3383219.3383247} \acmISBN{978-1-4503-7731-7/20/04}

\begin{document}  

\title{Quality Assessment of Online Automated Privacy Policy Generators: An Empirical Study}

\author{Ruoxi Sun}
\affiliation{%
    \institution{The University of Adelaide, Australia}
}
\email{ruoxi.sun@adelaide.edu.au}

\author{Minhui Xue}
\affiliation{%
    \institution{The University of Adelaide, Australia}
}
\email{jason.xue@adelaide.edu.au}

\renewcommand{\shortauthors}{Ruoxi Sun and Minhui Xue}

\begin{abstract}
Online Automated Privacy Policy Generators (APPGs) are tools used by app developers to quickly create app privacy policies which are required by privacy regulations to be incorporated to each mobile app. The creation of these tools brings convenience to app developers; however, the quality of these tools puts developers and stakeholders at legal risk. In this paper, we conduct an empirical study to assess the quality of online APPGs. We analyze the completeness of privacy policies, determine what categories and items should be covered in a complete privacy policy, and conduct APPG assessment with boilerplate apps. The results of assessment show that due to the lack of static or dynamic analysis of app's behavior, developers may encounter two types of issues caused by APPGs. First, the generated policies could be incomplete because they do not cover all the essential items required by a privacy policy. Second, some generated privacy policies contain unnecessary personal information collection or arbitrary commitments inconsistent with user input. Ultimately, the defects of APPGs may potentially lead to serious legal issues. We hope that the results and insights developed in this paper can motivate the healthy and ethical development of APPGs towards generating a more complete, accurate, and robust privacy policy.
\end{abstract}




\maketitle

%---------------------------------------------------
\section{Introduction}
%---------------------------------------------------

Mobile phones have become increasingly important to our daily lives. Information privacy issues are of growing concern to developers as well as to mobile app users. Nowadays, it is very common that a mobile app accesses, requires, or transmits users' sensitive personal information when requesting services. Privacy policy is one of the key legal statements that is used to protect users' personal information. Many countries have adopted privacy regulations, such as the Children's Online Privacy Protection Act (COPPA)~\cite{coppa}, Health Insurance Portability and Accountability Act (HIPPA)~\cite{HIPAA:02}, the European General Data Protection Regulation (GDPR)~\cite{EU:16}, and the California Online Privacy Protection Act (CalOPPA)~\cite{CalOPPA}. These regulations restrict the behavior of apps and define the obligations of app developers. One of the main requirements of these privacy regulations is that a privacy policy is mandatory for an app. For example, the Article 13.1(c) of GDPR requires that the use of a data subject's personal data should have a legal basis. The Federal Trade Commission (FTC) also recommended that app developers should ensure easy access to the privacy policy through app stores. 
However, not every app developer is a legal professional and writing privacy policies could be a time-consuming and tedious task~\cite{balebako2014improving}. The emergence of APPGs greatly eases such dilemma. For companies small in size or individual developers who cannot afford the expense of legal fees, an APPG online could be a good option, which can quickly create a customized privacy policy with basic or perhaps reliable terms. 

The ease of APPGs comes at a cost of legal risk. FTC severely penalizes companies for breach of commitments in their privacy policies~\cite{zimmeck2014privee}. Failing to provide a correct privacy policy may cause serious legal issues. For example, InMobi, a Singapore-based mobile advertising company, was fined \$950,000 by FTC as it collected user's location information without declaring the behaviors in the privacy policy~\cite{InMobi}. In 2014, a company of a popular flashlight app and its owner were complained by FTC as they did not reflect the app's use of personal data in app privacy policies. 
Users of the APPG may encounter the same legal issue with the APPG \textit{per se}.

In this paper, we make several key contributions listed below.
\begin{itemize}
\item 
    To assess privacy practices of online APPGs, we analyzed the completeness of privacy policies and determined what categories and items should be covered in a complete privacy policy.
\item 
    We investigated 10 popular online APPGs and then studied the complexity of use and payment methods used by them. We designed three boilerplate apps to generate privacy policies and used them for APPG quality assessment. Our method assesses the quality of APPGs by combining the aspects of completeness of privacy policy and the consistency with user input.
\item 
    Combined with the results of the quality assessment, we believe that the quality of free APPGs not significantly outperforms that of a well-designed template. We found that the quality of the paid APPGs largely outperforms the free APPGs, despite many functional issues arisen.
\end{itemize}

%---------------------------------------------------
\subsection{Related Work}\label{sec:related_work}
%---------------------------------------------------

\noindent \textbf{Privacy policy generation techniques.} Recent work has been proposed to generate privacy policies for apps. For instance, the AutoPPG~\cite{yu2016toward}  conducts static code analysis to extract app behaviors about user personal information. Based on the extracted behaviors, AutoPPG applies natural language processing (NLP) techniques to generate privacy policies.

\noindent \textbf{Mobile app privacy policy.} Several works have studied the consistency between privacy policies and app behaviors~\cite{harkous2018polisis, slavin2016toward}. Andow \etal~\cite{andow2019policylint} present a state-of-the-art privacy policy analysis tool, termed PolicyLint. Utilizing sentence-level NLP techniques, PolicyLint is able to detect the positive and negative statements in a privacy policy, which enables the automated analysis of the contradictions in a privacy policy. Chen \etal~\cite{chen2018ausera} analyzed the security risk of banking apps with NLP and static analysis. Zimmeck \etal~\cite{zimmeck2019maps} designed a scaling privacy compliance analysis tool through code analysis and machine learning techniques. 

\noindent \textbf{Our work.} Because the existing online APPGs rely heavily on user input and templates, if the quality of APPGs cannot be guaranteed, the privacy policy they generate may deviate from the terms expected by users, and inconsistencies with app behaviors may cause serious legal consequences. This paper differentiates from previous works by looking into the quality assessment of APPGs \textit{per se} in terms of their completeness and consistency.

%---------------------------------------------------
\section{Methodology}\label{sec:Methodology}
%---------------------------------------------------
\subsection{Collecting Online APPGs}

In order to investigate the current status of APPGs, we identified a set of popular app privacy policy generators. 
We first used Google search queries, including but not limited to  ``privacy terms template'', ``GDPR compliant privacy policy generator'', and ``mobile app privacy policy generator''. For each result, we selected the first ten entries, including advertising links provided by Google. We also collected recommendation of APPGs across forums, blogs~\cite{digital.com, oberlo.com}, or in the post of the legal consultant website~\cite{thelegality.com}. We remove the APPGs that cannot generate privacy policies for mobile apps and select the most popular ones in terms of their stated marketing scale and user recommendations. 

\begin{table*}[t]
    \scriptsize
    \centering
    \caption{Overview of online APPGs}
    \label{table:APPGs}
    \vspace{-3mm}
    \begin{tabular}{cp{4.0cm}llccc}
        \toprule
        \textbf{\#}&\textbf{APPG}& \textbf{Company}&\textbf{Location}&\textbf{Payment}&\multicolumn{1}{l}{\textbf{Has Privacy Policy?}} & \multicolumn{1}{l}{\textbf{Has Disclaimer?}} \\
        \midrule
        \textbf{1}&\textbf{seqlegal.com}                & SEQ Legal LLP & Oxfordshire, UK   & Free  & No    & No    \\
        \textbf{2}&\textbf{volusion.com}                & Volusion      & Austin, USA       & Free  & Yes   & No    \\
        \textbf{3}&\textbf{app-privacy-policy-generator.firebaseapp.com} & Individual  & N/A & Free  & No    & No    \\
        \textbf{4}&\textbf{getterms.io}                & Humaan        & Perth, Australia  & Free  & Yes   & Yes   \\
        \textbf{5}&\textbf{privacypolicygenerator.info}& N/A           & N/A               & Free  & No    & Yes   \\
        \textbf{6}&\textbf{legenova.com}               & Legal Nova Ltd& Sofia, Bulgaria   & Free  & Yes   & No    \\
        \textbf{7}&\textbf{termsfeed.com}   & N/A           & N/A               & Paid  & Yes   & No    \\
        \textbf{8}&\textbf{websitepolicies.com}  & WebsitePolicies Inc. & Toronto, Canada   & Paid & Yes   & Yes \\
        \textbf{9}&\textbf{wonder.legal}             & Miracle  & Paris, France     & Paid  & No    & No    \\
        \textbf{10}&\textbf{termly.io}                 & Termly LTD    & London, UK & Free/Period  & Yes   & Yes   \\
        \bottomrule
    \end{tabular}
\end{table*}

After removing duplicates and highly similar sites, we obtained top 10 popular APPGs listed in Table~\ref{table:APPGs}. Although most of them collect user Privacy Identifiable Information (PII), such as mailing address and email address, or even require financial information for service purchasing, only six of them have privacy policies to protect their customers' privacy. While APPGs having advertised their services are being widely used, 40\% of them also have disclaimers, stating that their services are only pieces of legal advice, instead of a formal legal service. For example, the APPG \#7 listed in Table~\ref{table:APPGs} claims that they provide ``Trusted legal agreement''. However, they also have a disclaimer stating that, ``TermsFeed.com is not a law firm and is not providing legal advice. All information (including agreements, forms and documents) available on our site, www.termsfeed.com, are provided without any warranty, express or implied''. Furthermore, for other APPGs, there is no company information provided, neither on their websites nor in payment receipts.

\begin{figure}[t]
    \centering
    \includegraphics[width=\linewidth]{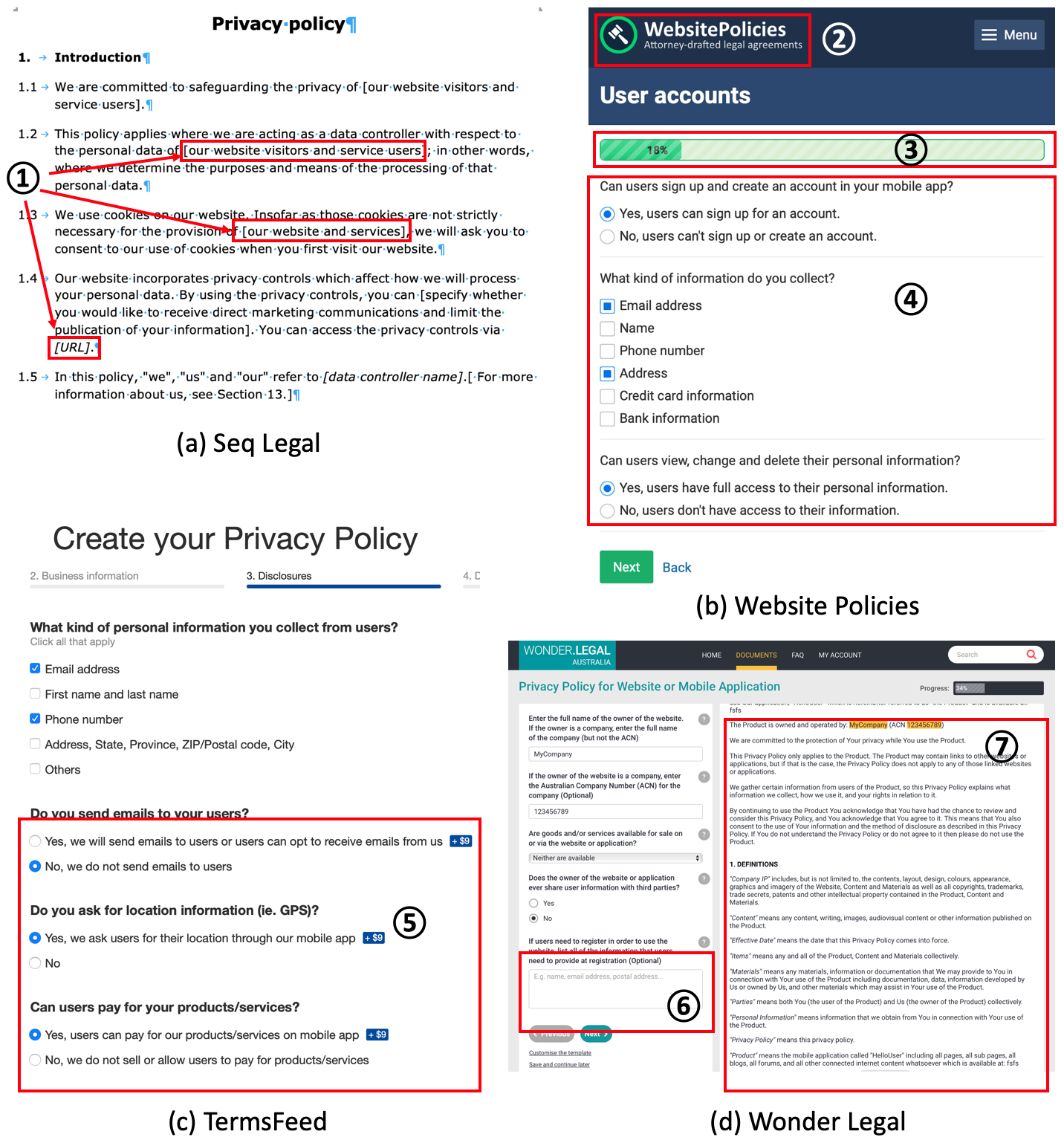}
    \caption{The UIs of APPGs. (1) Fill-in spaces in template. (2) Logo. (3) Progress bar. (4) Multiple choice questions. (5) Paid features. (6) Text area for user input. (7) Preview of policy.}
    \label{fig:APPG-UI}
    %\vspace{1mm}
\end{figure}

Figure~\ref{fig:APPG-UI} shows the user interfaces of typical APPGs. APPGs in a template form, \ie, \#1 and \#2, always have fill-in space for users to input detailed information. The user interface of survey questionnaire APPGs usually includes a logo, multiple choice questions with yes or no answers, and a progress bar. Some APPGs require payments to use advanced features. APPG \# 9 improves the flexibility of user input through an editable text area and provides a preview of the generated privacy policy.

More than half of the APPGs listed in Table~\ref{table:APPGs} are free of charge, or have a free version with limited functionalities, \eg, no GDPR compliant or can only generate one privacy policy. For the three paid APPGs, the user will be charged from \$20 to \$40 per privacy policy generated, or paid per item added into policy, \eg, \$14 for business using, \$9 for terms related to location information collection, \$14 for involving Google Analytics, which can make one privacy policy costs exceed \$180.

%---------------------------------------------------
\subsection{Completeness Analysis of Privacy Policies}
%---------------------------------------------------

As required in GDPR and other privacy regulations, apps must have a privacy policy to explicitly inform what kind of data is collected from users and how data is handled. Related works~\cite{guntamukkala2015machine, tesfay2018privacyguide}  also argue that a complete app privacy policy should include the following categories:

\begin{itemize}
    \item 
    \pb{The basic information,} which includes the identity of data collector, \eg, the name of the company, email address, and location of the company, the name of the app, and the date of version.
    
    \item 
    \pb{The personal information to be collected,} which includes PII and non-PII. As per the guide of PII protection by McCallister~\cite{mccallister2010guide}, the PII includes the name of a user, email address, IP address, location information, \etc However, due to the constant evolution of technologies, PII and non-PII may change and overlap.  %as per technological development~\cite{schwartz2011pii}. 
    Thus, we also consider non-PII information collected by apps, \eg, the content created by users in apps, device information, and information collected from any third parties.
    
    \item 
    \pb{Why and how the data are to be used,} which includes the purpose of data access, the identity of third parties, and whether the data will be disclosed as a response to legal requests. 
    
    \item
    \pb{Information management,} which includes the description of user rights, data transfers, the practice of data protection, the retention of data, the notification of change to privacy policies, whether the data will be shared in the case of  business merges and acquisitions, and asset sales, and the protection of children's data.  
\end{itemize}

Several most popular apps, \eg, Facebook, Google Calendar, Pokémon Go, Clash of Clans, Tik-tok, EBay, and Runtastic, were analyzed in this paper. Based on the data from Google Play Store, the downloads of these apps have all exceeded 50 million. Considering the spread scale and number of users of these apps, we can infer that their privacy policies are of high quality and minor defects. Through the completeness analysis on their privacy policies, we further investigated the categories that should be covered in privacy policies and determined essential items in each category as shown in Table~\ref{table:items}.

\begin{comment}
\begin{table} [ht]
    \footnotesize
    \centering
    \caption{Popular Apps}
    \label{table:apps}
    \vspace{-3mm}
    \begin{tabular}{llllr}
        \toprule
        \textbf{\#} & \textbf{APP} & \textbf{Type} & \textbf{Company} & \multicolumn{1}{l}{\textbf{Downloads}} \\
        \midrule
        \textbf{A}  & \textbf{Facebook}        & Social        & Facebook         & 5,000,000,000+                            \\
        \textbf{B}  & \textbf{Pokémon Go}      & Game          & Niantic          & 100,000,000+                              \\
        \textbf{C}  & \textbf{Google Calendar} & Business      & Google           & 1,000,000,000+                            \\
        \textbf{D}  & \textbf{Clash of Clans}  & Game          & Supercall        & 500,000,000+                              \\
        \textbf{E}  & \textbf{Tik-tok}         & Social        & Bytedance        & 500,000,000+                              \\
        \textbf{F}  & \textbf{eBay}            & Shopping      & eBay             & 100,000,000+                              \\
        \textbf{G}  & \textbf{Runtastic}       & Health        & Adidas           & 50,000,000+                              \\
        \bottomrule
    \end{tabular}
\end{table}
\end{comment}

\begin{table}[t]
    \footnotesize
    \centering
    %\vspace{-4mm}
    \caption{Essential categories and items in privacy policies}
    \label{table:items}
    \vspace{-3mm}
    \begin{tabular}{p{2cm}p{5.5cm}}
        \toprule
        \textbf{Categories}           & \textbf{Items}    \\
        \midrule
        \textbf{Basic information} & Name of company, Address of company, Name of app, Contact info, and Date of last version.   \\
        \midrule
        \textbf{Privacy Information Collected} & Name of user, Email address, Phone number, Username/Account ID, IP address, Location information, Purchasing information,  Content created by user, Information from third-parties, Device information, User activity, and Cookies. \\
        \midrule
        \textbf{Use of Information}            & For detecting abuse and illegal activity, For safety and security, For services improving, For service providing, Personalised advertising, Business analysis, Communication with users, Sharing with other users, Sharing with third-party, and Respond to legal requests. \\
        \midrule
        \textbf{Information Management}        & User rights, Data transfers, Protection of data, Data retention, Changes notifying, Business Transactions, Age. and Limits/Children's privacy  \\
        \bottomrule
    \end{tabular}
     %\vspace{-3mm}
\end{table}

In our investigation, two APPGs (\#1 and \#2) are in form of templates. All the remaining APPGs have a survey questionnaire, querying questions from users, and generate privacy policies based on the user input. Table~\ref{table:competeness_result} shows the result of APPG questionnaires. The symbol~\pie{360} indicates that the item is covered by the questionnaire. As we can see, most APPGs are covered no more than half of the essential items, indicating two potential defects: 1) for some essential items, users cannot input any information to the generator; 2) even if the generated privacy policies cover more items than its questionnaire does, the descriptions of items are not able to closely follow user input. For example, APPG \#4 does not even request which privacy information will be collected, but directly writes, ``we only ask for personal information when we truly need it to provide a service to you. We collect it by fair and lawful means, with your knowledge and consent,'' in the generated privacy policy.

We also counted how many questions in each survey questionnaire of APPGs. Two template-typed APPGs have few questions; however, the most complex APPG has 45 questions for users to answer, and it may take roughly an hour to generate one privacy policy. 

\begin{table}[t]
    \scriptsize
    \centering
    \caption{Categories and items covered in APPG questionnaires (\# APPG is referred to the first column of  Table~\ref{table:APPGs})}
    \label{table:competeness_result}
    \vspace{-3mm}
    \begin{tabular}{l|cccccccccc}
        \toprule
        \textbf{\# APPG }                  & \textbf{1} & \textbf{2} & \textbf{3} & \textbf{4} & \textbf{5} & \textbf{6} & \textbf{7} & \textbf{8} & \textbf{9} & \textbf{10} \\
        \midrule
        \textbf{Basic information}                 &            &            &            &            &            &            &            &            &            &             \\
        Name of company                   &            & \pie{360}          & \pie{360}          & \pie{360}          & \pie{360}          &            & \pie{360}          & \pie{360}          & \pie{360}          & \pie{360}           \\
        Address of company                &            &            &            & \pie{360}          &            &            & \pie{360}          & \pie{360}          &            & \pie{360}           \\
        Name of App                       &            &            &            & \pie{360}          & \pie{360}          & \pie{360}          & \pie{360}          & \pie{360}          & \pie{360}          &             \\
        Contact info                      &            & \pie{360}          &            &            & \pie{360}          & \pie{360}          & \pie{360}          & \pie{360}          &            & \pie{360}           \\
        Date of last version              &            &            &            & \pie{360}          &            &            &            &            & \pie{360}          & \pie{360}           \\
        \midrule
        \textbf{Privacy Information Collected}     &            &            &            &            &            &            &            &            &            &             \\
        Name of user                      &            &            &            &            &            &            & \pie{360}          & \pie{360}          &            &             \\
        Email Address                     &            &            &            &            &            & \pie{360}          & \pie{360}          & \pie{360}          &            &             \\
        Phone                             &            &            &            &            &            &            & \pie{360}          & \pie{360}          &            &             \\
        Username/Account ID               &            &            &            &            &            &            &            &            &            &             \\
        IP address                        &            &            &            &            &            &            &            &            &            &             \\
        Location information              &            &            &            &            &            & \pie{360}          & \pie{360}          &            &            & \pie{360}           \\
        Purchase information              &            &            &            &            &            &            &            & \pie{360}          &            &             \\
        In-App sharing                    &            &            &            &            &            &            &            &            &            &             \\
        Content created by user           &            &            &            &            &            &            &            &            &            &             \\
        Information from third-parties    &            &            &            &            &            &            &            &            &            &             \\
        Device information                &            &            &            &            &            &            &            &            &            &             \\
        User activity                     &            &            &            &            &            &            &            &            &            &             \\
        Cookie data                       &            &            & \pie{360}          &            &            &            &            &            &            &             \\
        Others                            &            &            &            &            & \pie{360}          & \pie{360}          & \pie{360}          &            & \pie{360}          & \pie{360}           \\
        \midrule
        \textbf{Use of Information}                &            &            &            &            &            &            &            &            &            &             \\
        Detect abuse and illegal activity &            &            &            &            &            &            & \pie{360}          &            &            &             \\
        Safety and security               &            &            &            &            &            &            &            &            &            &             \\
        Improve products/services         &            &            &            &            &            &            &            &            &            &             \\
        For service providing             &            &            &            &            &            &            &            &            &            &             \\
        Personalised advertising          &            &            & \pie{360}          &            &            &            & \pie{360}          & \pie{360}          & \pie{360}          & \pie{360}           \\
        Business analysis                 &            &            &            &            &            & \pie{360}          & \pie{360}          & \pie{360}          & \pie{360}          &             \\
        Communication with users          &            &            &            &            &            &            & \pie{360}          & \pie{360}          & \pie{360}          & \pie{360}           \\
        Sharing of information            &            &            &            &            &            &            &            &            &            &             \\
        Sharing with other users          &            &            &            &            &            & \pie{360}          &            & \pie{360}          & \pie{360}          & \pie{360}           \\
        Sharing with third-party          &            &            &            &            &            &            &            & \pie{360}          & \pie{360}          &             \\
        Respond to legal requests         &            &            &            &            &            &            & \pie{360}          & \pie{360}          &            &             \\
        \midrule
        \textbf{Information Management}            &            &            &            &            &            &            &            &            &            &             \\
        User rights                       &            &            &            &            &            &            &            & \pie{360}          &            &             \\
        Data transfers                    &            &            &            &            &            &            &            &            &            &             \\
        Protection of data                &            &            &            &            &            &            &            &            & \pie{360}          & \pie{360}           \\
        Data retention                    &            &            &            &            &            &            &            &            &            & \pie{360}           \\
        Changes notifying                 &            &            &            &            &            &            &            & \pie{360}          &            &             \\
        Business Transactions             &            &            &            &            &            &            & \pie{360}          & \pie{360}          & \pie{360}          &             \\
        Age limits/Children's privacy     &            &            &            &            &            &            &            & \pie{360}          & \pie{360}          & \pie{360}           \\
        \midrule
        \textbf{Number of Questions}               & 0          & 2          & 4          & 4          & 6          & 13         & 20         & 30         & 45         & 45   \\
        \bottomrule
    \end{tabular}
\end{table}

%---------------------------------------------------
\subsection{Designing Boilerplate Apps}
%---------------------------------------------------

To further assess the quality of APPGs, we designed boilerplate apps to simulate the privacy policy generation process. For each APPG, we generated privacy policies based on the wording of the three same boilerplate apps to evaluate whether the generated privacy policy is covered by essential items and whether the generated terms are aligned with user input.

\begin{itemize}
    \item
    \textbf{App~1:} The first app only collects the full name of the user which is visible on the screen. This app will only disclose user information in the case of legal request.
    
    \item
    \textbf{App~2:} A game app utilizing augmented reality technology collects user's location information. Users need to register an account to login the game app, which collects user’s name and email address. Users are able to login with a third-party account, such as a Google account or a Facebook account. Users can purchase virtual items in-app through a third-party payment processor. The design of this app comes from the popular game Pokémon Go.
    
    \item
    \textbf{App~3:} The third app is an social app which enables users interaction and communication with each other. Users can create contents and share them publicly in-app. Users need to register an account to login the social app. The registration process collects user’s personal information, such as name, email address, phone number. The app also utilizes cookies data for updating services. 
\end{itemize}

Table~\ref{table:dummy apps} lists all the categories and items that could be incorporated in the generated privacy policies. We defined a three-level criticality of items in an app: \textit{Not related, Recommended, and Essential}. \textit{Not related} means that the item is unnecessary to be explicitly mentioned in the privacy policy. For example, App~1 does not exploit user information for safety and security purpose, regardless of the reference, we give the corresponding item $0$ score; \textit{Recommended} means that the item is not critical; however, an incorrect description may cause potential inconsistency and legal issues; \textit{Essential} items should be exactly described in the privacy policy as per user input and inconsistent description may cause serious legal issues. For example, App~3 collects email address from users, and this must be described consistently in the generated privacy policy. 

\begin{table} [t]
    \small
    \centering
    \caption{Design of dummy apps}
    \label{table:dummy apps}
    \vspace{-3mm}
    \begin{tabular}{lccc}
    \toprule
    \textbf{Identity of Data Collector}        & \textbf{App 1}             & \textbf{App 2}             & \textbf{App 3}             \\
    \midrule
    Name of company                   & \pie{180} & \pie{180} & \pie{180} \\
    Address of company                & \pie{180} & \pie{180} & \pie{180} \\
    Name of App                       & \pie{180} & \pie{180} & \pie{180} \\
    Contact info                      & \pie{180} & \pie{180} & \pie{180} \\
    Date of last version              & \pie{180} & \pie{180} & \pie{180} \\
    \midrule
    \textbf{Information Collection}   & \textbf{App 1}             & \textbf{App 2}             & \textbf{App 3}             \\
    \midrule
    Name of user                      & \pie{360} & \pie{360} & \pie{360} \\
    Email Address                     & \pie{180} & \pie{360} & \pie{360} \\
    Phone                             & \pie{180} & \pie{180} & \pie{360} \\
    Username/Account ID               & \pie{180} & \pie{360} & \pie{360} \\
    IP address                        & \pie{180} & \pie{360} & \pie{360} \\
    Location information              & \pie{180} & \pie{360} & \pie{360} \\
    Purchase information              & \pie{180} & \pie{360} & \pie{360} \\
    In-App sharing                    & \pie{180} & \pie{180} & \pie{360} \\
    Content created by user           & \pie{180} & \pie{180} & \pie{360} \\
    Information from third-parties    & \pie{180} & \pie{360} & \pie{360} \\
    Device information                & \pie{180} & \pie{180} & \pie{360} \\
    User activity                     & \pie{180} & \pie{180} & \pie{360} \\
    Cookie data                       & \pie{180} & \pie{180} & \pie{360} \\
    Others                            & \pie{180} & \pie{180} & \pie{360} \\
    \midrule
    \textbf{Use of Information}       & \textbf{App 1}             & \textbf{App 2}             & \textbf{App 3}             \\
    \midrule
    Detect abuse and illegal activity & \pie{90}   & \pie{360} & \pie{360} \\
    Safety and security               & \pie{90}   & \pie{90}   & \pie{360} \\
    Improve products/services         & \pie{90}   & \pie{360} & \pie{360} \\
    For service providing             & \pie{360} & \pie{360} & \pie{360} \\
    Personalised advertising          & \pie{180} & \pie{180} & \pie{360} \\
    Business analysis                 & \pie{180} & \pie{180} & \pie{360} \\
    Communication with users          & \pie{180} & \pie{360} & \pie{360} \\
    \midrule
    \textbf{Information Disclosure}   & \textbf{App 1}             & \textbf{App 2}             & \textbf{App 3}             \\
    \midrule
    Sharing with other users          & \pie{360} & \pie{360} & \pie{360} \\
    Sharing with third-party          & \pie{360} & \pie{360} & \pie{360} \\
    Respond to legal requests         & \pie{360} & \pie{360} & \pie{360} \\
    \midrule
    \textbf{Information Management}   & \textbf{App 1}             & \textbf{App 2}             & \textbf{App 3}             \\
    \midrule
    User rights                       & \pie{180} & \pie{180} & \pie{180} \\
    Data transfers                    & \pie{180} & \pie{180} & \pie{180} \\
    Protection of data                & \pie{180} & \pie{180} & \pie{180} \\
    Data retention                    & \pie{180} & \pie{180} & \pie{180} \\
    Changes notifying                 & \pie{180} & \pie{180} & \pie{180} \\
    Business Transactions             & \pie{180} & \pie{180} & \pie{180} \\
    Age limits/Children's privacy     & \pie{180} & \pie{180} & \pie{180} \\
    \bottomrule
    \end{tabular}
    \\\pie{90}: Not related~~~\pie{180}: Recommended~~~\pie{360}: Essential
\end{table}

%---------------------------------------------------
\subsection{Quality Assessment of APPGs}
%---------------------------------------------------

Based on the criticality level of items for each boilerplate app, we evaluate the quality of generated privacy policies. The rules of evaluation are described in Table~\ref{table: rules}. If the description exactly matches the user input, we will give a high score to the item; otherwise, a negative penalty score will be given. Considering there are many descriptions generated based on templates, which may be still correct but not exactly be consistent with the user input, we will assign a positive but a lower score. For example, whatever the user input is for a policy generated by APPG \#2, it always has the sentence ``we value your trust in providing us your Personal Information, thus we are striving to use commercially acceptable means of protecting it.'' We will only give score $1$ for such blur description as this may lead to potential inconsistency and increase risks to developers and app users.

Using this method, we cannot only evaluate the completeness of a generated privacy policy, but also assess the APPG quality based on whether the terms match the user input. A higher score indicates better performance of an APPG and a lower score indicates that there are more inconsistencies with user inputs or more terms generated by templates.

\begin{table} [t]
    \footnotesize
    \centering
    \caption{The rules of APPG performance evaluation}
    \label{table: rules}
    \vspace{-3mm}
    \begin{tabular}{llc}
    \toprule
    \textbf{Criticality of Items}               & \textbf{Consistency with User Input} & \textbf{Score} \\
    \midrule
    \textbf{Not related}                  & N/A                         & 0     \\
    \textbf{Essential}   & Exactly consistent          & 5     \\
                                 & Not exactly consistent      & 3     \\
                                 & Inconsistent                & -5    \\
    \textbf{Recommended} & Exactly consistent          & 3     \\
                                 & Not exactly consistent      & 1     \\
                                 & Inconsistent                & -3   \\
    \bottomrule
    \end{tabular}
\end{table}

We also measure the similarity between generated privacy policies and the original privacy policies of boilerplate apps, \ie, privacy policies from Facebook and Niantic.\footnote{Niantic (https://nianticlabs.com) is a world's leading augmented reality game company, the developer of Pokémon Go.} We first store privacy policies into text files, then convert all uppercase characters into lowercase and remove all punctuation marks. We compute the TF-IDF~\cite{rajaraman2011mining} values for each word in privacy policies, where the word weighting is encoded by Equation~\eqref{eq:1} as follows.

\begin{equation}
    \label{eq:1}
    w_{i,j} = {tf}_{i,j} \cdot \log \left(\frac{N}{df_i}\right),
\end{equation}
where $w_{i,j}$ is the weight for word $i$ in policy $j$, $tf_{i,j}$ is the frequency of word $i$ in policy $j$, $N$ is the number of documents in the collection, and $df_i$ is the document frequency of word $i$ in the collection. After applying TF-IDF analysis, we obtained a weight vector for each word in the privacy policy, where a high value indicates a high weight of this word in that policy. 

Cosine similarity is often used as an indicator of similarity between two documents. We applied this method according to Equation~\eqref{eq:2} to measure the similarity between the generated privacy policy and its original policy. 

\begin{equation}
    \label{eq:2}
    \text{cosine~similarity} = \frac{V_{g} \cdot V_{o}}{\left\lVert V_{g}\right\rVert \cdot \left\lVert V_{o}\right\rVert}, 
\end{equation}
where $V_{g}$ is the TF-IDF weight vector of generated privacy policy, $V_{o}$ is the weight vector of original privacy policy. Because we design the functionalities of the boilerplate apps based on the original apps, a highly weighted word in the original privacy policy should also be weighted highly in the generated privacy policy. Hence, a high similarity rate indicates a high-quality APPG.

%---------------------------------------------------
\section{Results and Analysis}\label{sec:result analysis}
%---------------------------------------------------

\begin{figure}[t]
    \centering
    \includegraphics[width=0.95\linewidth]{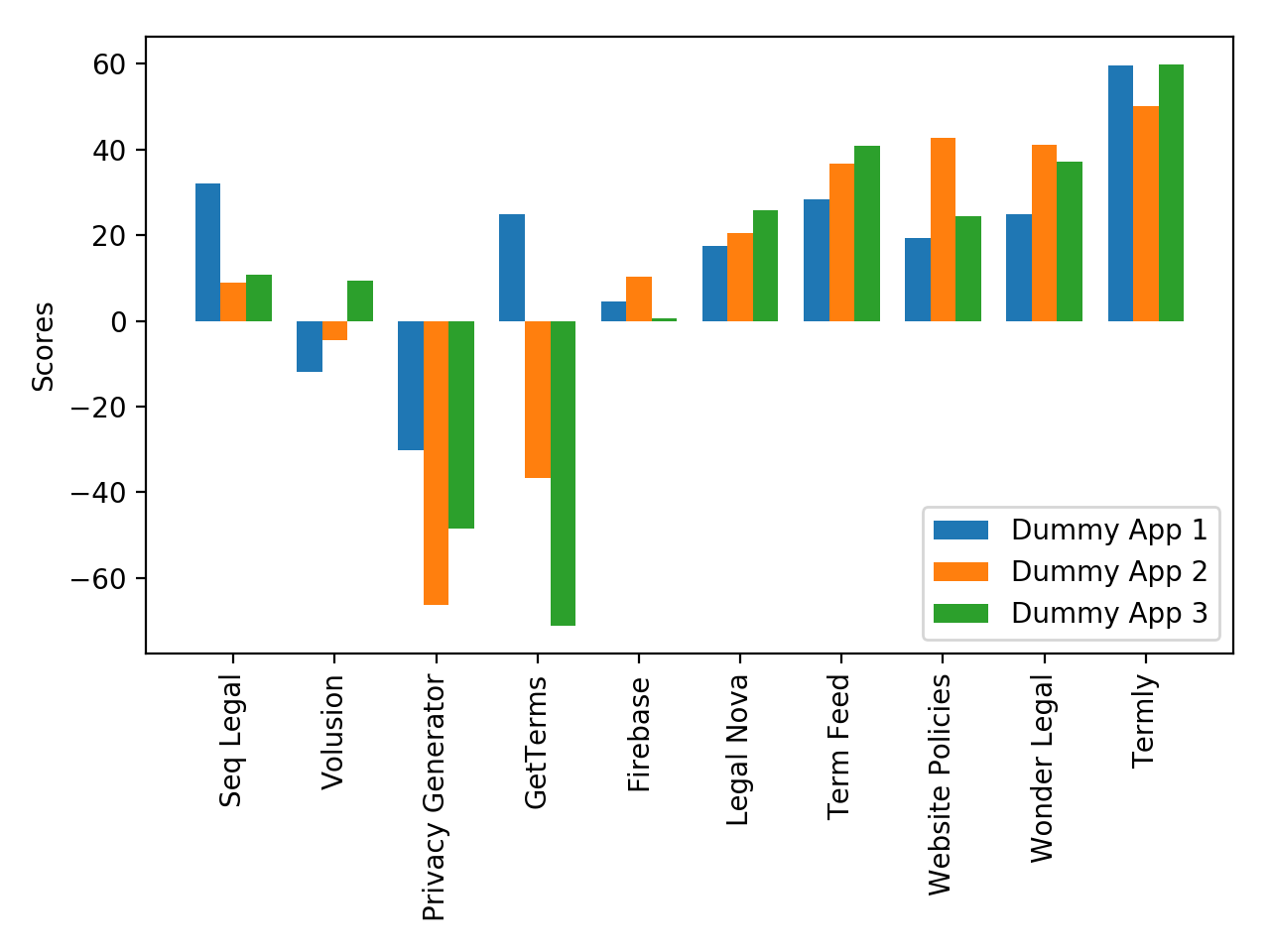}
    \vspace{-3mm}
    \caption{The performance scores of APPGs evaluated by dummy apps.}
    \label{fig:2_scores_dummy_apps}
    %\vspace{-3mm}
\end{figure}

To analyze the quality of APPGs, we generated 30 privacy policies with 3 boilerplate apps and 10 APPGs.  Figure~\ref{fig:2_scores_dummy_apps} shows the quality assessment results of these privacy policies. The scores are scaled into $[-100, 100]$ for comparison. The t-test applied on free and paid app scores indicates that there is significant difference in the performance of free and paid APPGs (see findings in detail below). 

\noindent \textbf{Finding 1.} \textit{The versatility of template is limited.} We can see that, for APPGs \#1--\#4, the assessment results of the three boilerplate apps are different, because they are in the form of templates or rely heavily on templates to generate privacy policies. The functions of apps are different from each other, so their scores change drastically. One template may generate a good quality privacy policy for a specific app, but it may perform poorly when being applied to another app. For example, APPG \#1 always generates a term that the app will not collect personal information from third parties. This is good enough for App~1, but it fails to match App~2 and App~3. The result shows that it is not a good solution to apply one template to all apps. 

\noindent \textbf{Finding 2.} \textit{Some APPGs generate over-scoped terms.} For APPGs \#6, \#7, and \#9, the scores generally increase with the complexity of the boilerplate apps. This is mainly because the APPGs generate some personal information terms which are over-scoped of the ground truth that user inputs. For example, whatever the user input is, APPG \#6 will, by default, generate terms to collect personal information, \eg, email address, IP address, location information, which is heavily penalized for App~1 in our quality assessment. 

\begin{figure}[t]
    \centering
    \includegraphics[width=0.94\linewidth]{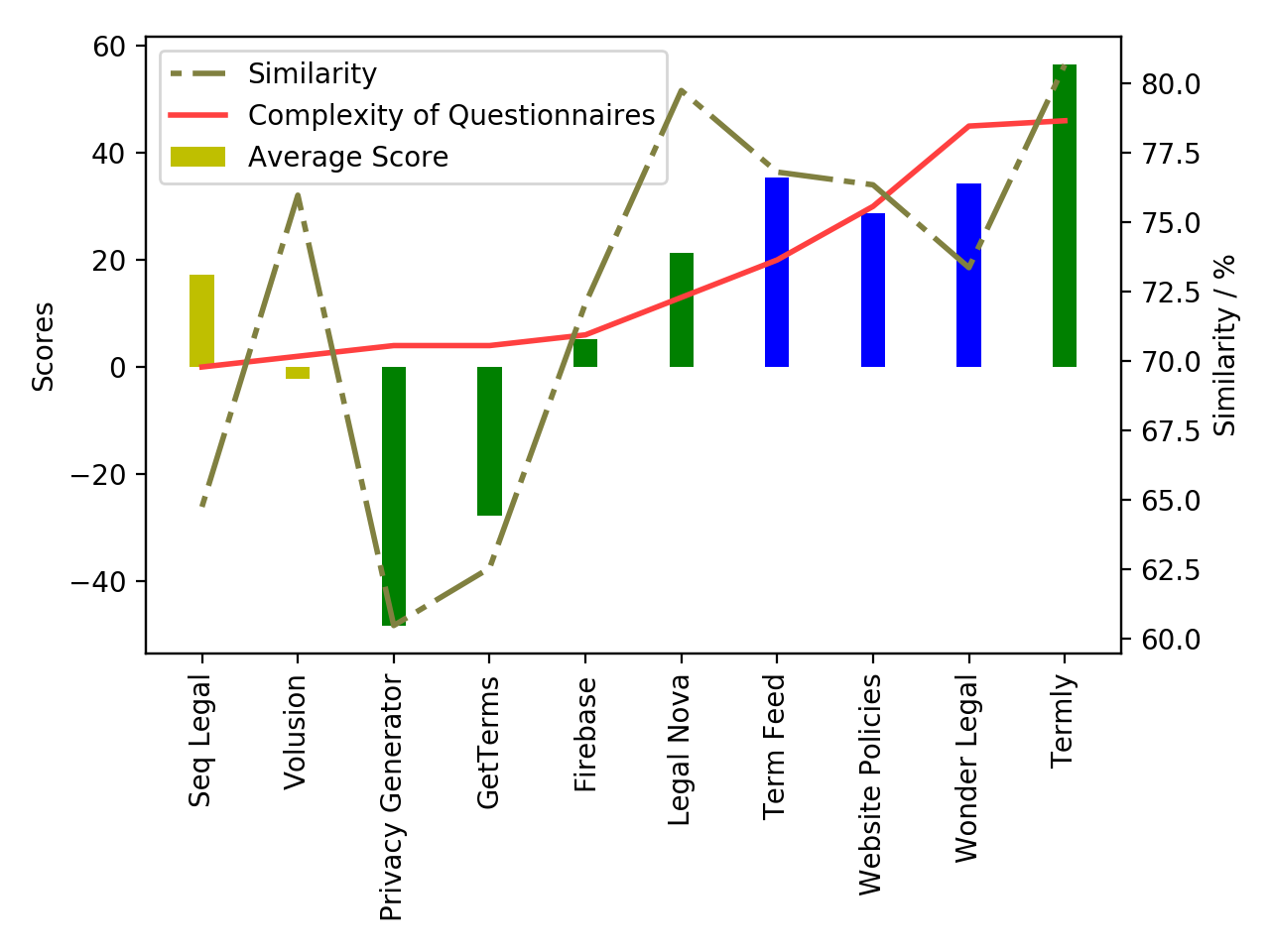}
    \vspace{-4mm}
    \caption{The performance of APPGs vs. complexity of questionnaire and similarity to original policies.}
    \label{fig:complexity_similairity_scores}
    \vspace{-1mm}
\end{figure}

As shown in Figure~\ref{fig:complexity_similairity_scores}, we calculated the average quality score of each APPG and measure the similarity between generated privacy policies and the original ones. The scores of templates, free, and paid APPGs are marked as yellow, green, and blue, respectively. The red line represents the complexity of questionnaires encoded by the number of questions. The dashed line represents the similarity results. 

\noindent \textbf{Finding 3.} \textit{Keeping generated terms consistent with user input and ensuring the coverage rate of essential items is critical to high-quality policy generation.} The performance of APPGs mostly increases with the complexity of the questionnaire. However, for APPGs \#8 and \#9, despite more complex questionnaires, their performance is lower than APPG \#7, which indicates that more complex survey questionnaires do not necessarily perform better. To achieve high quality, APPGs should consider more  essential items in the design of questionnaires so as to ensure that a complete privacy policy is generated. In addition, users should be allowed to answer questions in a more flexible way, as some questions require users to provide more specific information rather than simply answering multiple choice questions; otherwise, the terms generated may not be consistent with the actual behavior of the app. For example, although APPG \#8 has 30 questions in the survey, it never covers the item of cookies access and there is no text area for users to input which personal information will be collected. It will always claim the collection of device's IP address, location, and device name, lowering its quality score.

\noindent  \textbf{Finding 4.} \textit{The similarity score fluctuates with the quality assessment outcome.} The dashed line of Figure~\ref{fig:complexity_similairity_scores} reflects the similarity result of APPGs. For several APPGs, such as \#1, \#2, and \#6, there are some gaps between similarity trend and scoring due to the fact that more penalties are pushed towards the low-quality terms which do not fit into actual user input.

\noindent \textbf{Finding 5.} \textit{APPG quality is positively correlated to payment.} Considering the payment mode of APPGs shown in Table~\ref{table:APPGs}, free APPGs, such as \#3--\#6, perform less commendably than a well-designed template, evidenced by the fact that their quality scores are no greater than the template APPG \#1. APPG \#10 is an exception as it needs to be paid periodically for retrieving advanced features. Meanwhile, paid APPGs,such as \#7--\#9, normally have a more complex questionnaire and better perform than free APPGs. 

%---------------------------------------------------
\section{Conclusion}\label{conclusion}
%---------------------------------------------------

We analyzed the privacy policies of popular apps and specified what categories and items a complete privacy policy should cover. Based on the essential items of privacy policy, we provide a method to assess the quality of online APPGs. The assessment results show that the performance of APPGs largely increases with the complexity of the survey questionnaires. Although there are still many template-generated terms which are potentially inconsistent with user's input, paid APPGs normally outperform free APPGs as a whole. 

Most existing APPGs are in form of survey questionnaires. Even some APPGs claim that they do not make any templates, the method they applied is to collect user's answers and to fit them into a ``smart" template. They only generate privacy policies relied on user's inputs (\eg, answer questions). As lack of static or dynamic code analysis of apps, while using APPGs, developers and app users may encounter two types of serious issues. First, the generated policies could be incomplete because they did not cover all the essential items required by a privacy policy, \eg, did not include all personal information collected by the app. Second, some generated policies may be over-scoped, \eg, they claim to collect personal information which is not in fact  accessed by the app, and this may lead to potential inconsistency and legal issues. We strongly recommend that developers should systematically consult a professional legal team before releasing a privacy policy generated by APPGs. In future work, besides extending assessment to more APPGs, we will further optimize the APPG quality assessment through semantic analysis, and will conduct a more in-depth assessment as per the requirements of a specific regulation, \eg, HIPAA or GDPR.

\bibliographystyle{ACM-Reference-Format}
\bibliography{paper}

\end{document}